# Huge Thermoelectric Power Factor: FeSb$_2$ versus FeAs$_2$ and RuSb$_2$

Peijie Sun, Niels Oeschler, Simon Johnsen[1], Bo B. Iversen[1], and Frank Steglich

Max Planck Institute for Chemical Physics of Solids, D-01187 Dresden, Germany

[1]Department of Chemistry, University of Aarhus, DK-8000 Aarhus C, Denmark

The thermoelectric power factor of the narrow-gap semiconductor FeSb$_2$ is greatly enhanced in comparison to the isostructural homologues FeAs$_2$ and RuSb$_2$. Comparative studies of magnetic and thermodynamic properties provide evidence that the narrow and correlated bands as well as the associated enhanced thermoelectricity are only specific to FeSb$_2$. Our results point to the potential of FeSb$_2$ for practical thermoelectric application at cryogenic temperatures and stimulate the search for new correlated semiconductors along the same lines.

A narrow distribution of electronic density of states (DOS) close to the Fermi level has long been suggested to generate a high thermoelectric (TE) performance [1]. This sort of electronic structure emerges in many *f*- or *d*-electron based materials at low temperatures. Nevertheless, efforts thus far have not yielded any promising compound for low-*T* (*T* < 77 K) application [2]. A high thermoelectric dimensionless figure of merit *ZT* (= $TS^2/\rho\kappa$, with *S*, $\rho$ and $\kappa$ being thermoelectric power, electrical resistivity, and thermal conductivity, respectively) is achieved by either enhancing the so-called power factor *PF* = $S^2/\rho$, through, e.g., optimization of electronic structure [1], and/or reduction of $\kappa$. Prevalent TE materials are heavily doped semiconductors with energy gap $E_g$ > 0.1 eV [2]. Such an $E_g$ in classic semiconductors only guarantees a high *ZT* value around or above 200 K, whereas decreasing $E_g$ induces a rapid compensation of *S* between electrons and holes, thus a low *ZT*.

In this connection, the recent finding of a huge *PF* in FeSb$_2$ [3] near 10 K provides experimental evidence for the theoretical expectation: The group of semiconductors with narrow energy gaps and correlated bands could be good thermoelectrics at cryogenic temperatures. Like the well-studied *d*-based correlated semiconductor FeSi [4], FeSb$_2$ is fascinating for fundamental study of electron-electron correlations as well [5-7]. The maximum absolute value of *S*, |$S_{max}$|, in FeSb$_2$ amounts to 45 mV/K, the largest reported for compound semiconductors. The *ZT* is low, exclusively due to the large $\kappa$ values. Nevertheless, FeSb$_2$ appears to be attractive for TE exploration because i) the enhanced *PF* is principally electronic in origin, as was suggested by Bentien *et al.* [3] and was recently proved by us [8] and, ii) its $\kappa$ is overwhelmingly dominated by the lattice contribution below 200 K. That is to say, *PF* and $\kappa$ are decoupled in FeSb$_2$, which provides a basis for optimization of the *ZT* value by reducing $\kappa$ selectively. This work compares FeSb$_2$ to isoelectronic and isostructural RuSb$_2$ and FeAs$_2$, the latter two systems bearing no (or slight) indications of electron correlations.

Here, an extra contribution to the thermoelectric power solely present in $FeSb_2$ is highlighted. The absence of this part in the two reference compounds stresses the relevance of electron-electron correlations to the enhanced thermoelectricity in $FeSb_2$, clearly pointing to its potential for low-$T$ application.

Single crystals of $FeSb_2$ prepared by self-flux (SF) method as used in previous work [3] and new ones prepared by chemical vapor transport (CVT) technique were employed. The samples synthesized by the CVT technique were grown by utilizing $Br_2$ as transport agency and keeping the feed zone and growth zone at 700 and 650°C, respectively. The reference compounds $RuSb_2$ and $FeAs_2$ were prepared by SF and CVT, respectively. X-ray diffraction analysis shows that they indeed crystallize in the common marcasite structure. The high purity of the samples was verified by the small carrier concentration down to $10^{21}$ m$^{-3}$, as well as the marked increase of $\kappa$ at low $T$ characteristic of phonon-phonon umklapp scattering. After orienting the samples with Laue back-reflection technique, their TE properties were measured by a commercial physical properties measurement system (PPMS, Quantum Design) as described in ref. 3. The thermopower and thermal conductivity of the newly prepared $FeSb_2$ were further cross-checked on a home-made cryostat with Au(Fe0.07%)-Chromel thermal couple as temperature gradient detector. The typical sample dimension for the transport measurements is $(4.0 - 5.0) \times 2.0 \times 0.4$ mm$^3$ for $FeSb_2$, and the length is even longer (5-8 mm) for $RuSb_2$ and $FeAs_2$.

All three compounds have ever been studied in search of new TE materials for high-$T$ application [9-11]. Band-structure calculations by different authors show a similar energy gap in their respective DOS at Fermi level, i.e., 0.1 - 0.3 eV in $FeSb_2$ [12-14], 0.3 eV in $FeAs_2$ [12] and 0.2 eV in $RuSb_2$ [14]. For $FeSb_2$, this is a direct gap and was detected by optical conductivity ($E_g^{dir}$ = 0.13 eV) [15]. However, the optical spectral weight suppressed due to the gap opening is not recovered just above the gap, but is extended to above 1 eV [6], indicative of strong electron-electron correlations. Resistivity measurements for various $FeSb_2$ samples give two smaller gaps, $E_{g1}$ = 4 - 10 meV and $E_{g2}$ = 26 - 36 meV (cf. inset of Fig. 1 and ref. 3), which are presumably in-direct gaps, as were also suggested by band-structure calculation [13] and optical conductivity [15]. For $FeAs_2$ and $RuSb_2$, on the other hand, thermally activated $\rho(T)$ near room temperature indicates an energy gap of 0.20 and 0.29 eV, respectively (inset of Fig. 1). This is again in rough agreement to the band calculations. Unlike $FeSb_2$, the latter two compounds exhibit high-resistance metallic behavior above 50 K, which presumably marks impurity conduction as is usually observed in classic semiconductors. The smaller transport gaps $E_{g1}$ and $E_{g2}$ in $FeSb_2$, however, are strongly believed to be intrinsic: They are independent of sample quality and reproducible in many physical properties besides $\rho$, including optical conductivity [15], Nernst effect [8], and specific heat as will be shown below.

The comparison of the thermoelectric power $S$ among them reveals significant differences (Fig. 1). While the maximum thermopower values are found at roughly the same temperature, 10 K, their values are very different. $|S_{max}|$ of $FeSb_2$ is an order of magnitude larger than that of the other two. The position of $|S_{max}|$ coincides with the minimum in the Hall coefficient and is therefore, determined by the thermally activated charge carriers. Classical theories can predict the carrier-diffusion part of $S$ for a semiconductor by three carrier parameters: concentration $n$, effective mass $m^*$, and scattering rate [2]. Our attempts to interpret such a large difference in $|S_{max}|$ solely based on $m^*$ was unfruitful, considering the similar $n$ for all samples [8]. Another important contribution to $S$ in semiconductors (particularly, elemental semiconductors like germanium and silicon [16]) is the phonon-drag effect. This effect appears to be evident when the corresponding $n$ is low and $\kappa$ is high, with a large phonon mean free path. A large $\kappa$ is indeed observed in the currently studied systems (inset of Fig. 2), to which the enhanced $|S_{max}|$ can partially be attributed. Nevertheless, in view of the same crystal structure and the even larger $\kappa$ of FeAs$_2$ and RuSb$_2$, it seems natural to exclude phonon-drag effect as the *dominating* contribution and attribute the additional, huge contribution to $|S_{max}|$ of FeSb$_2$ to its correlated bands. As discussed below, magnetic and thermodynamic properties will add further evidence in favor of this argument.

The enhanced $|S_{max}|$ below 20 K leads to a huge power factor *PF* for FeSb$_2$ (Fig. 2). The maximum *PF* exceeds 2000 µW/K$^2$cm at around 10 K, whereas it is below 200 µW/K$^2$cm for FeAs$_2$ and RuSb$_2$. The huge *PF* in FeSb$_2$ dwarfs the prevalent values in optimized Bi$_2$Te$_3$, 40 µW/K$^2$cm, and the largest *PF* known in YbAgCu$_4$, 235 µW/K$^2$cm [2]. With regard to the enhanced values of *PF* in FeSb$_2$, the high carrier mobility (yielding a relatively small $\rho$) [17] should be stressed in addition to the huge $|S_{max}|$. This large *PF* value, combined with a reduced $\kappa$ of ~1 W/Km, which is achievable in view of the minimum $\kappa$ prediction [18], indicates a practical *ZT* value larger than unity.

An intuitive demonstration of the $|S_{max}|$ enhancement in FeSb$_2$ is illustrated by the so-called Jonker plot [19] (Fig. 3), which depicts carrier-concentration dependence of thermoelectric power at a certain temperature (here 10 K, the position of $|S_{max}|$). The carrier concentration *n* is deduced from the Hall coefficient (data partly shown in refs.8 and 20) by means of a one-band model. Apparently, $|S_{max}|$ of FeSb$_2$ exceeds that of FeAs$_2$ or RuSb$_2$ by one order of magnitude at the lowest *n*. Even when *n* increases tenfold in FeSb$_2$, its $|S_{max}|$ is still substantially larger than that of its counterparts. It is worth noting that the $|S_{max}|$ - log *n* dependence for FeSb$_2$ has a slope more than 30 times larger than the prediction for a classical, noncorrelated semiconductor [19].

Though an enhanced *S* is indeed anticipated for FeSb$_2$ due to its narrow, enhanced energy bands, the microscopic origin of the enhanced |*S$_{max}$*| and *PF* in FeSb$_2$, which exceed that of FeSi by two orders of magnitude, is not yet clear. A microscopic description of the thermoelectricity for FeSb$_2$ should involve a proper treatment of electron-electron correlations at the borderline of the degenerate and nondegenerate regimes [8], and merge the unique magnetic and thermodynamic properties of FeSb$_2$ shown below (Fig. 4). The magnetic susceptibility $\chi$(*T*) of FeSb$_2$ displays thermally activated, enhanced Pauli paramagnetism above 50 K, whereas FeAs$_2$ and RuSb$_2$ exhibit an almost *T*-independent diamagnetism, ascribed to their inner core contribution. Enhanced Pauli paramagnetism is an imprint of the enhanced *m\** that is expected for FeSb$_2$. Actually, $\chi$(*T*) of FeAs$_2$ slightly increases above 200 K, too. In view of its moderately enhanced *PF*, we do not exclude the possibility that FeAs$_2$ could resemble FeSb$_2$ to some degree, regarding electron-electron correlations. On the other hand, the difference of the specific heat between FeSb$_2$ and RuSb$_2$, $\triangle C$ (= $C_{FeSb2} - C_{RuSb2}$), exhibits two peaks at around 50 and 200 K. This extra specific heat results from the thermal population of charge carriers between the narrow bands in FeSb$_2$. Here RuSb$_2$ was used as a reference compound, because its Debye temperatures $\theta_D$ is similar to that of FeSb$_2$ ($\theta_D$ is 348 and 359 K for FeSb$_2$ and RuSb$_2$, respectively, in contrast to a much higher value in FeAs$_2$, 510 K). Fitting the Schottky-type anomalies to a double-interval energy scheme by the formula $\Delta C = A \frac{(\Delta_1/k_BT)^2 \exp \Delta_1/k_BT}{(1+\exp \Delta_1/k_BT)^2} + B \frac{(\Delta_2/k_BT)^2 \exp \Delta_2/k_BT}{(1+\exp \Delta_2/k_BT)^2}$, we find two energy scales, $\Delta_1$ = 11.2 meV, and $\Delta_2$ = 50.9 meV. Good agreement with the above-mentioned transport gaps strongly indicates that they are intrinsic features of FeSb$_2$. The intrinsic in-direct gaps (or in-gap states) and the strong electron-electron correlations, which are associated with the enhanced thermoelectricity in FeSb$_2$, too, distinguish FeSb$_2$ from its two counterparts.

In conclusion, we compared the thermoelectric properties of the isostructural semiconductors $FeSb_2$, $FeAs_2$, and $RuSb_2$. The outstandingly enhanced thermoelectricity in $FeSb_2$ indicates a dominating contribution from its unique band structure and correlated charge carriers. The correlation-enhanced thermoelectricity is of particular interest for practical thermoelectric application in the cryogenic temperature range. Introducing nanometer scale internal structures to selectively scatter propagating phonons appears to be promising for creating a sufficiently high *ZT* value.

**Figure Captions**

FIG.1. Thermoelectric power $S$ for $FeSb_2$, $FeAs_2$, and $RuSb_2$, which all have a similar carrier concentration below 30 K (cf. Fig. 3, too). Inset shows their electrical resistivity as $\rho$ vs $T$. Data for $FeSb_2$ are taken from reference [3]. Electrical/heat current was applied along the $c$-axis for $FeSb_2$ and $RuSb_2$, and along the $a$-axis for $FeAs_2$. Thermal activation behavior in $\rho$ in certain $T$ ranges, with $\rho = \rho_0 \exp E_g/2k_BT$, yields $E_{g1}$ = 4.5 meV ($T$: 5 - 15 K) and $E_{g2}$ = 31.1 meV ($T$: 50 - 200 K) for $FeSb_2$, $E_g$ = 0.20 eV ($T$ > 200 K) for $FeAs_2$ and $E_g$ = 0.29 eV for $RuSb_2$ ($T$ > 250 K).

FIG. 2. The thermoelectric power factor $PF$ (main part) and the thermal conductivity $\kappa$ (inset) for the same samples shown in Fig. 1.

FIG. 3. $S_{max}$ vs carrier concentration $n$ for $FeSb_2$, $FeAs_2$, and $RuSb_2$ at $T \approx$ 10 K. The letters adjacent to the points denote the crystal directions along which the transport properties were measured. Additional points that are not described in the text are also included. The solid line (green) denotes the prediction for nondegenerate free electrons, and the dashed line (red) indicates a slope 35 times that of the prediction line.

FIG. 4. Magnetic susceptibility $\chi$ for $FeSb_2$, $FeAs_2$, and $RuSb_2$ (left axis) and the electronic specific-heat divided by $T$, $\Delta C/T$, for $FeSb_2$ (right axis). Solid line represents a fit to $\Delta C/T$ based on a double-interval energy scheme.

FIG.1

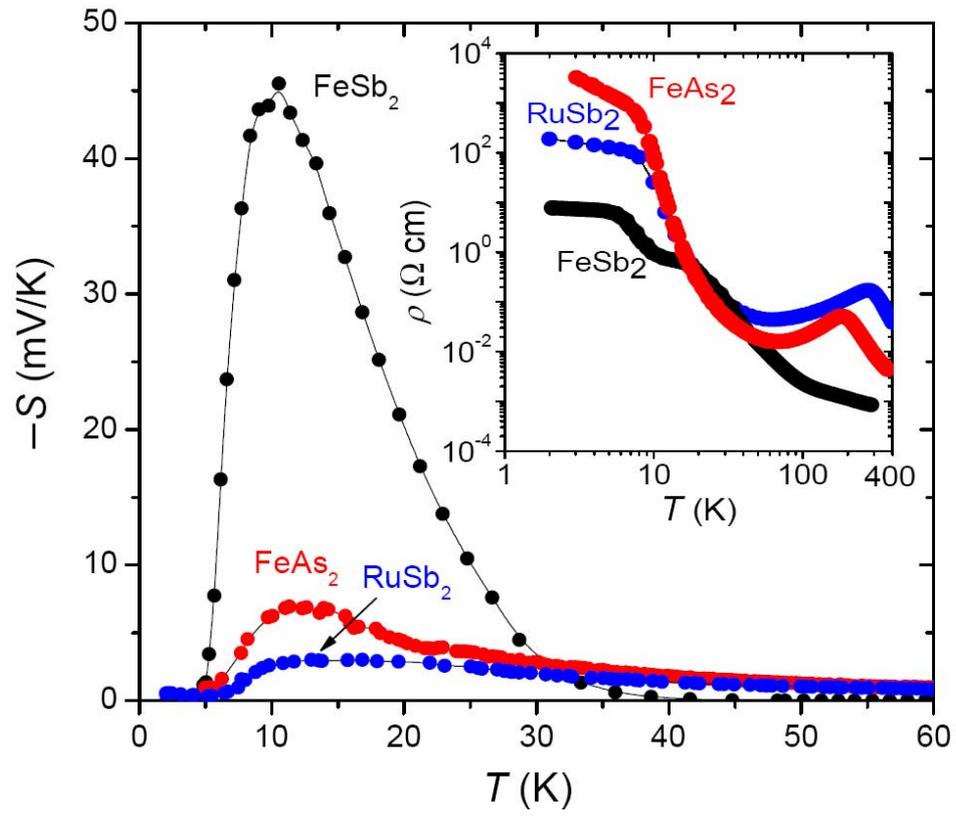

FIG.2

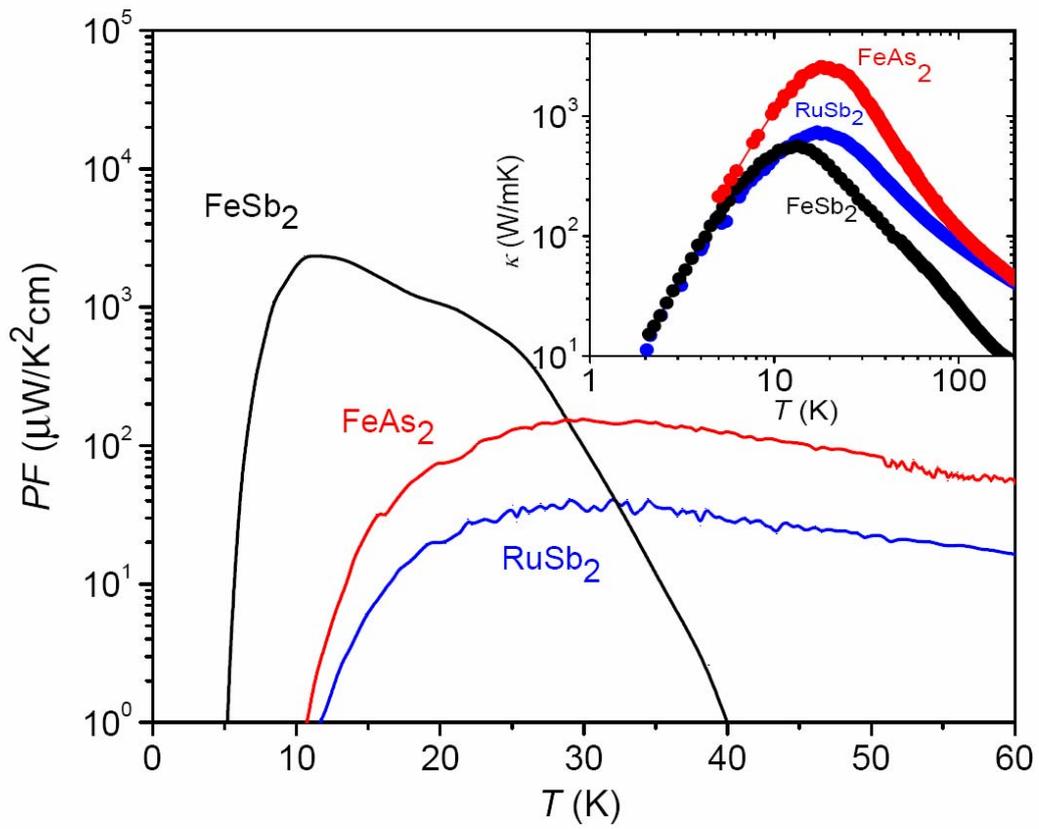

FIG.3

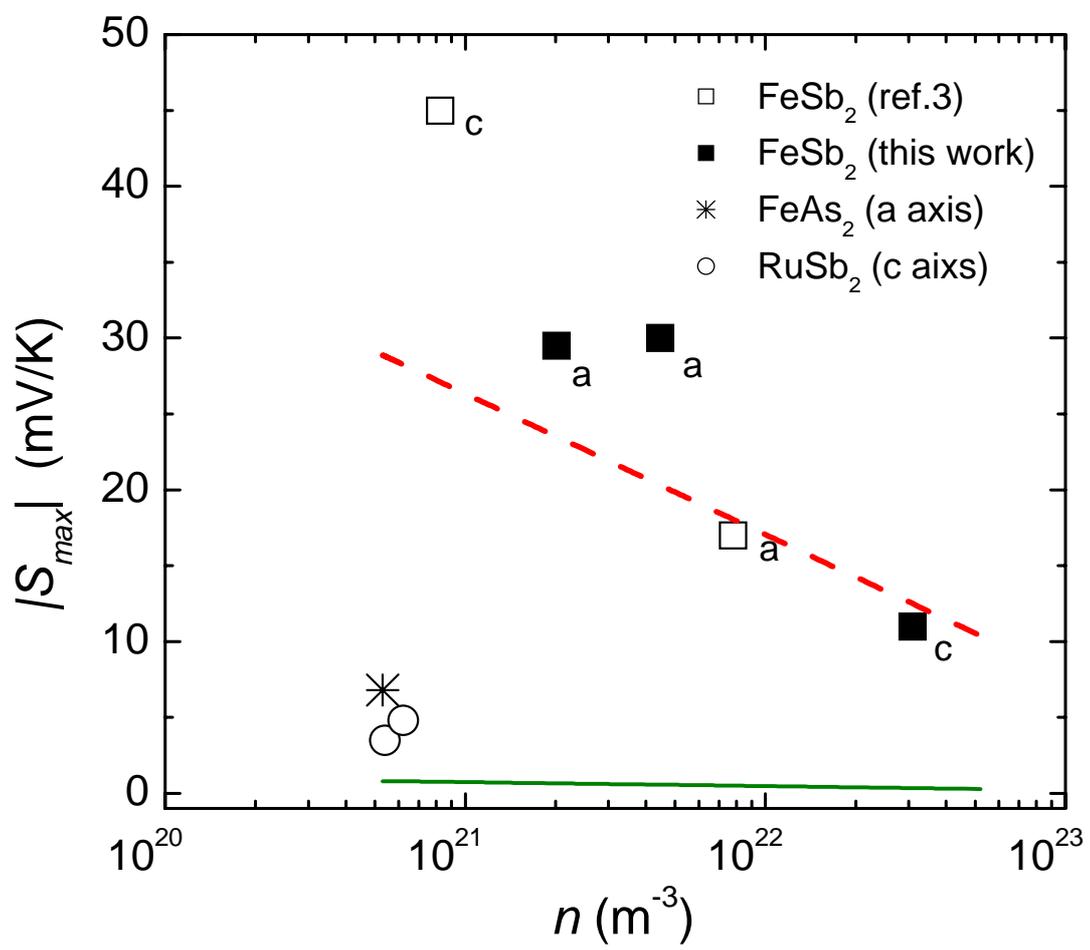

FIG.4

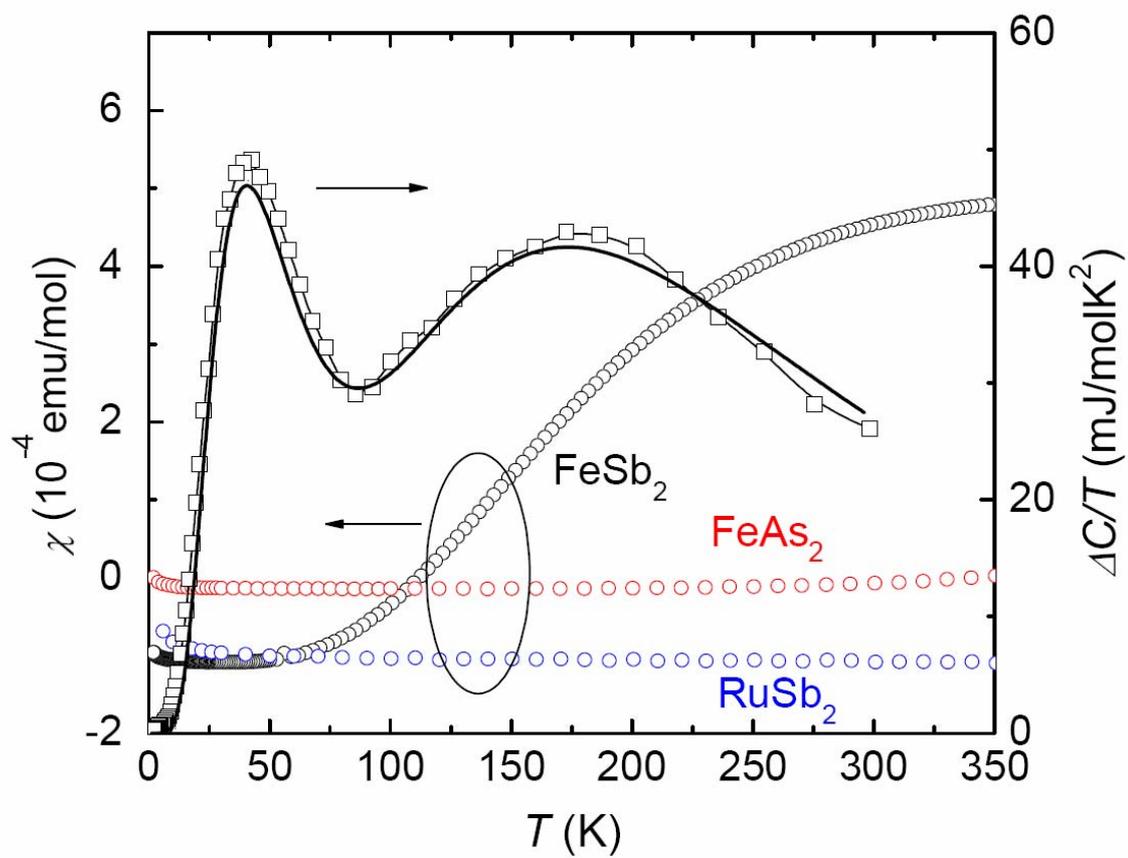